\documentclass[12pt,draftclsnofoot,journal,onecolumn]{IEEEtran}
\IEEEoverridecommandlockouts
\usepackage{algorithm}
\usepackage{amsmath}
\usepackage{algpseudocode}
\usepackage{graphicx}
\usepackage{wasysym}

\usepackage{mathtools}
\usepackage{enumitem} 
\usepackage{ifpdf}
\usepackage{amsthm}
\usepackage{amsmath}
\usepackage{nicefrac}
\usepackage{cite}
\usepackage{amsfonts}
\usepackage{comment}
 \usepackage{url}
 \usepackage{amssymb}
 \usepackage[paper=letterpaper,margin=0.75in]{geometry}
 \usepackage[ansinew]{inputenc}

\usepackage{lipsum}
\usepackage{hyperref}
\theoremstyle{definition}



\begin{document}
\title{Privacy-preserving mHealth Data Release with Pattern Consistency}

\author{
	\IEEEauthorblockN{Mohammad Hadian$^1$, Xiaohui Liang$^1$, Thamer Altuwaiyan$^1$, and Mohamed M E A Mahmoud$^2$}
	
	\IEEEauthorblockA{$^1$Department of Computer Science, University of Massachusetts Boston\\
	}
	\IEEEauthorblockA{$^2$Department of Electrical and Engineering, Tennessee Technological University\\
	}
	Email: \{mshadian,xiaohui,thamerfa\}@cs.umb.edu; mmahmoud@tntech.edu
	    
                   \thanks{ This work has been presented at the IEEE Global Communications Conference (GLOBECOM), 2016}
                   \thanks{\copyright Personal use of this material is permitted. Permission from IEEE must be obtained for all other uses, in any current or future media, including reprinting/republishing this material for advertising or promotional purposes, creating new collective works, for resale or redistribution to servers or lists, or reuse of any copyrighted component of this work in other works.}

}

\date{}
\maketitle
\thispagestyle{plain}
\pagestyle{plain}

\newtheorem{definition}{Definition}

\begin{abstract}
Mobile healthcare system integrating wearable sensing and wireless communication technologies continuously monitors the users' health status. However, the mHealth system raises a severe privacy concern as the data it collects are private information, such as heart rate and blood pressure. In this paper, we propose an efficient and privacy-preserving mHealth data release approach for the statistic data with the objectives to preserve the unique patterns in the original data bins. The proposed approach adopts the bucket partition algorithm and the differential privacy algorithm for privacy preservation. A customized bucket partition algorithm is proposed to combine the database value bins into buckets according to certain conditions and parameters such that the patterns are preserved. The differential privacy algorithm is then applied to the buckets to prevent an attacker from being able to identify the small changes at the original data. We prove that the proposed approach achieves differential privacy. We also show the accuracy of the proposed approach through extensive simulations on real data. Real experiments show that our partitioning algorithm outperforms the state-of-the-art in preserving the patterns of the original data by a factor of $1.75$.
\end{abstract}

\section{Introduction}
%Electronic Health (eHealth) and Mobile Health (mHealth) mechanisms are recently introduced to replace traditional paper based health record systems.
The mobile healthcare (mHealth) system with emerging wearable devices and wireless communications has removed geographical and distance related barriers of healthcare and has made the continuously-collected health data available anytime and anywhere~\cite{istepanian2006m,liang2012enable,lu2013spoc,hersh2002medical}. However, the mHealth system raises a severe privacy concern to its users because the mHealth data usually contain private information~\cite{dong2011challenges}, such as heart rate and blood pressure, at a highly fine-grained level, which may be maliciously used to derive patients' sensitive information, such as daily activities and health conditions. Keeping privacy of the data is the most important challenge of the emerging mHealth system~\cite{lu2013spoc,sahama2013security}.

\begin{figure}[hb]
	\centering
	\includegraphics[width=.4\textwidth]{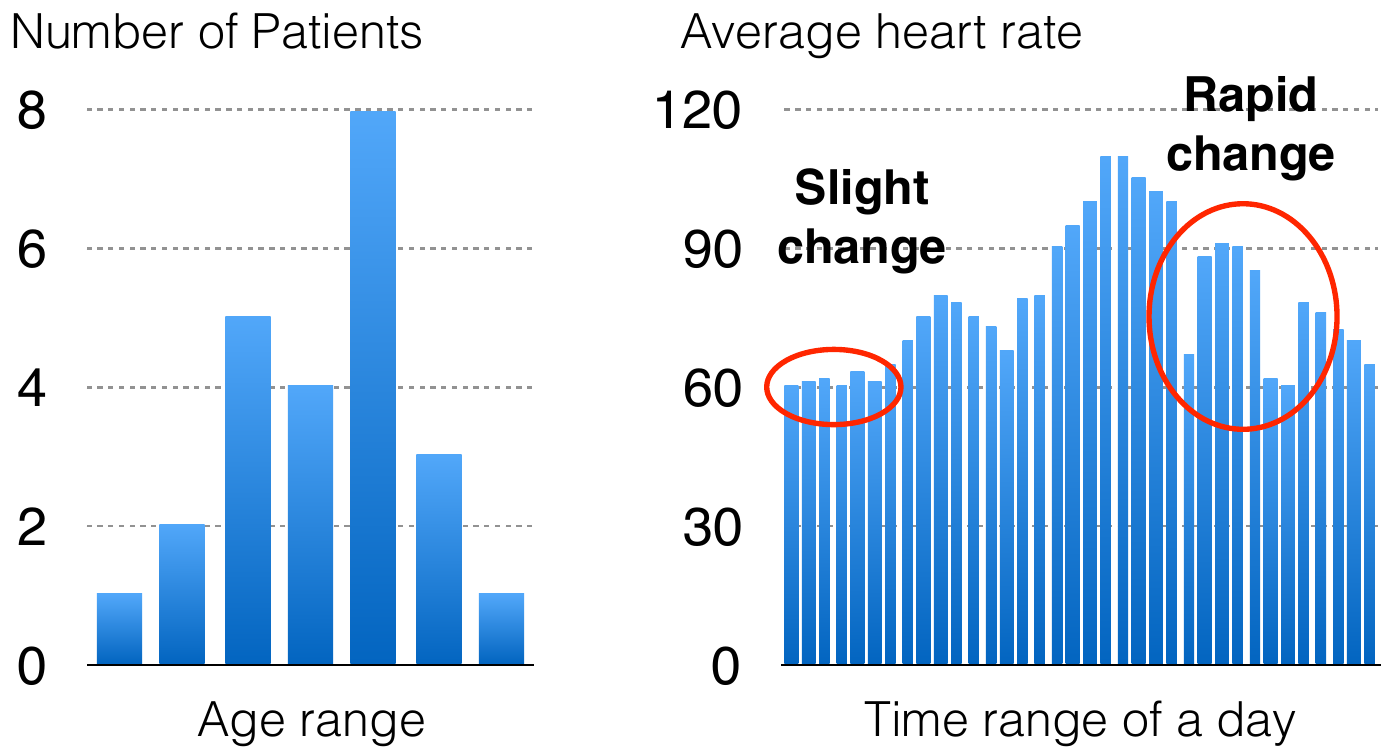}
	\vspace{-0.2cm}
	\caption{Histogram and mHealth data bins}\label{diff}
	\vspace{-0.2cm}
\end{figure}

Differential Privacy (DP) \cite{DBLP:reference/crypt/Dwork11} has been studied to help release data with privacy protection and accuracy assurance. DP can simply be explained as the mechanism of randomizing the results of the given query by adding some noise, commonly generated through Laplace distribution, to the original results based on a privacy budget $\epsilon$ to preserve the privacy of the individual records in the database. DP can guarantee mathematically-provable and measurable privacy preservation because of its precise definition and proof of privacy protection \cite{dwork2006calibrating}. DP has been applied to release the data of the histogram in Fig. \ref{diff} (left) \cite{li2015ehealth, li2014data}. With DP, it can be guaranteed that an attacker cannot distinguish the original histogram and the histogram with any bin plus 1 or minus 1. However, the aggregate mHealth data are much different from histogram bins from three perspectives (shown in Fig. \ref{diff}): i) The bin values usually represent real values but not counts; ii) The bins can be generated at a high frequency, such as heart rate being continuously monitored by wearable devices at a rate of one per minute; and iii) The bins may be interpreted with special diagnosis needs and some particular pattern consistency is required. All these distinctive features make the existing DP mechanisms~\cite{xu2013differentially, li2014data} inefficient when applied to the mHealth data.

%The other important advantage of these mechanisms is to make the data available for research through data release.

In this paper, we focus on an efficient and privacy-preserving data release approach specially designed for the mHealth data. We observe that the pattern consistency is particularly important for the mHealth data analysis. If the pattern of the original data is removed for the privacy concern, we consider such privacy-preserving approach is unacceptable in the mHealth data analysis. Here, we particularly study two pattern consistency requirements: i) if the difference of the values of two bins are larger than a threshold, the relationship of the outputs corresponding to the two bins must be preserved; and ii) if the difference of the values of two adjacent bins are larger than a threshold, the relationship of the outputs must be preserved. Naturally, the difference threshold for the adjacent bins is smaller than the first one to distinguish between rapid and slight changes. The motivation behind these requirements is to enable healthcare providers to observe the patients' health conditions from the preserved patterns. Specifically, we propose a novel differential privacy mechanism to release the aggregate query results of the mHealth data as shown in Fig. \ref{diff} (right). To preserve the defined pattern, the DP mechanism consists of two stages: the first stage is the private partitioning of the database bins into DP-compliant buckets, and the second stage is the noise addition to the average value in each bucket for data release to achieve DP. % in which we use the notion of relative error in order to achieve higher accuracy.
%exploit the data-dependent noise generation mechanism for gaining more accuracy through relative error as the second stage.
Specifically, the contributions of this paper can be summarized as follows:

$\bullet$ First, we study the patterns occurring in the mHealth data and observe that there are two distinguishable patterns of ``rapid change" and ``slight change" in these data, which the current privacy preserving techniques are unable to distinguish and preserve them in the final result. The slight change is referred to the minor changes that happen over the time, while the rapid changes happen between two adjacent data bins.

$\bullet$ Second, we propose an efficient and privacy-preserving mHealth data release mechanism. A new pattern-preserving partitioning algorithm is developed to preserve the rapid changes from the slight changes in the released data.

$\bullet$ Third, we prove that the proposed mechanism achieves $\epsilon-$differential privacy. We further obtain the accuracy analysis results through an extensive simulations using real data set. The results show that our algorithm achieves pattern consistency while other existing mechanisms cannot.

%The remainder of this paper is organized as follows. In Section II, we describe the preliminaries. In Section III, we introduce the details of our model and algorithm. After that we give a strict privacy proof of our algorithm followed by the evaluation in Section IV and Section V, respectively. Then we mention related works in Section VI and finally concludes the paper in Section VII.


\section{Preliminaries}

\subsection{Histogram and privacy}

Histograms, are commonly used to aggregate information to represent statistical data, as shown in Fig. \ref{diff} (left). The statistical data usually are considered to be privacy-preserving because the attacker is unable to derive the information about a single data record. However, if the database allows the attacker to run multiple queries on histograms without any constraints, the attacker is able to find out a newly-added or deleted data record by simply running the same query before and after the operation. For example, in Fig. \ref{diff} (left), the attacker queries the number of patients for all the age ranges of $n$ and $(n+1)$ consecutive days. Then, the attacker is able to derive the age of a patient admitted on $(n+1)$-th day, which is not intended to be disclosed.

\subsection{Differential privacy}

Differential privacy (DP) mechanisms are recently proposed to address the above privacy problem. The idea is to tolerate a small change in the database and prevent the attacker from being able to tell the change. Here we detail the DP definitions.

DP places a bound (controlled by privacy budget $\epsilon$) on the difference in the probability of algorithm outputs for any two neighboring databases. For any given database instance $I$, let $nbrs(I)$ denote the set of neighboring databases which differ from $I$ in at most one record; i.e., if $I' \in nbrs(I)$, then $|(I - I') \cup (I' - I)| = 1$.
\begin{definition}
	A randomized algorithm $\mathcal{A}$ is $\epsilon-$Differentially Private if for all instances $I$, any $I' \in nbrs(I)$, and any subset of outputs $S \subseteq Range(\mathcal{A})$, the following holds:
	\begin{center}
		$Pr[\mathcal{A}(I) \in S] \leq exp(\epsilon) \times Pr[\mathcal{A}(I') \in S] $
	\end{center}
\end{definition}

\begin{definition}[Sensitivity of a query]
	Given a sequence of counting queries $Q$, the sensitivity of $Q$, denoted as $\Delta Q$, is:
	\begin{center}
		$\Delta Q = max ||Q(I)-Q(I')||_1$
	\end{center}
	where $I' \in nbrs(I)$.
\end{definition}

\textbf{Laplace mechanism:} We use $Lap(b)$ to denote the Laplace probability distribution with mean $0$ and scale $b$. The Laplace mechanism is commonly used to achieves DP by adding Laplace noise to a query output.
\begin{definition}
	Let $Q$ be a query sequence of length $d$ and $\mathcal{Z}$ be a $d$-length vector of random variables where $\mathcal{Z}_i \AC Lap(\Delta Q_i/\epsilon)$. The Laplace mechanism \~{Q} is defined as:
	
	\begin{center}
		\~{Q}$(I)=Q(I) + \mathcal{Z}$
	\end{center}
	The randomized algorithm \~{Q} is $\epsilon$-differentially private.
\end{definition}

Differential private algorithm has two properties.\\
$\bullet$ Sequential composition: If there are $n$ independent
algorithms $\mathcal{A}_1,. . . ,\mathcal{A}_n$, whose privacy budgets are $\epsilon_1, . . ., \epsilon_n$ respectively, any function $\mathcal{K}$ of them: $\mathcal{K}(\mathcal{A}_1,. . . ,\mathcal{A}_n)$ is
$\sum_{i=1}^n \epsilon_i$-differentially private.\\
$\bullet$ Parallel composition: If the previous mechanisms are computed on disjoint subsets of the private dataset then the function $\mathcal{K}$ would be $max\{\epsilon_i\}$-differentially private.
%For instance, in Fig.1 (left) we show a database containing blood sugar of patients collected from mobile devices along with time and certain blood sugar threshold for that person on that specific time. This information can be useful for doctors to follow the patterns of their patient's dining and activities and also other parties like insurance companies while selling insurance. 
%
%
%Keeping the pattern of the original data and reflecting it into the output is very important. For instance the blood sugar rates highly depend on the time of the day (e.g. mornings v.s. afternoons): high blood sugar in the mornings is a common sign in diabetic patients and also dropping the blood sugar after each meal does not happen in diabetic patients. Current DP algorithms which simply add Laplace noise to the original data fail to capture and preserve the trend in original data. Lost of these patterns in the process of DP may lead to wrong diagnosis.

\section{Proposed scheme}

In this section, we introduce the proposed privacy-preserving data release scheme. %We first introduce the four steps of the scheme, and then particularly detail the DP-compliant partitioning algorithm and randomizing algorithm.

\subsection{Scheme overview}

\begin{figure}[htbp]
	\includegraphics[width=.5\textwidth]{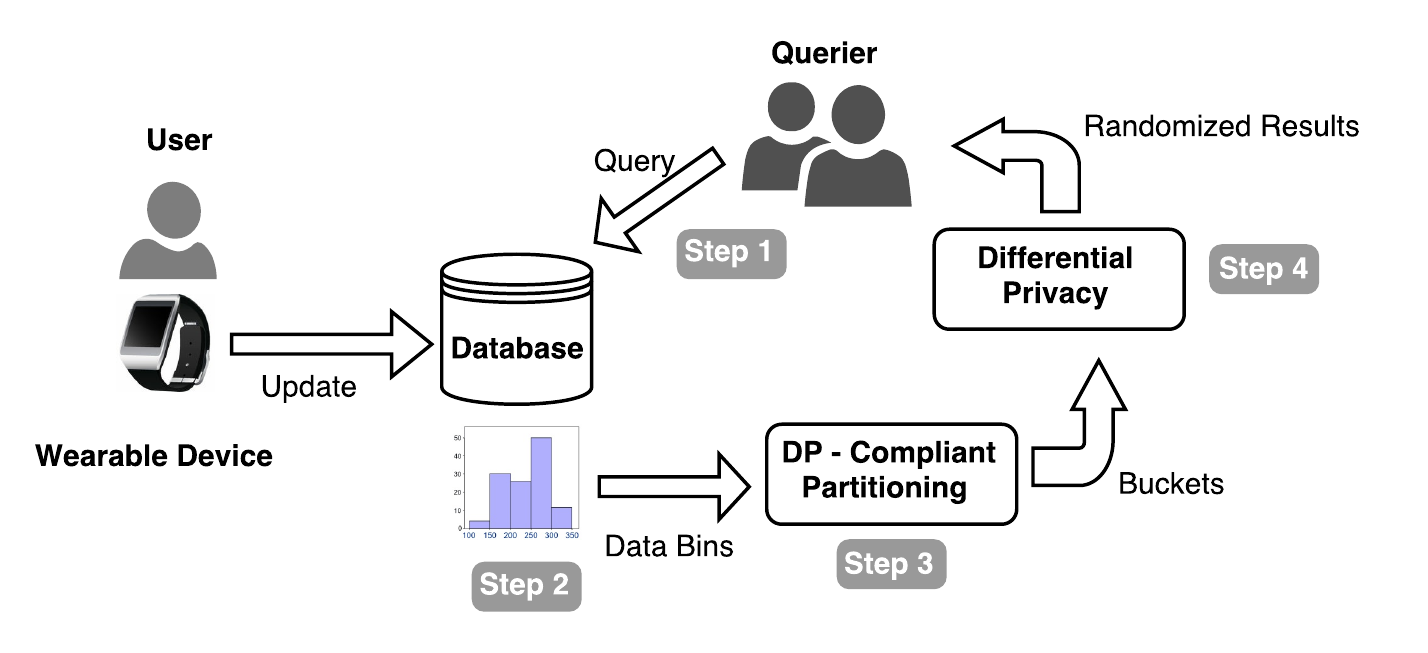}
	\vspace{-0.5cm}
	\caption{System Model}
	\vspace{-0.4cm}	
	\label{model}
\end{figure}

The overview of our scheme is shown in Fig. \ref{model}. Consider a scenario where the user has a wearable watch to monitor her health data continuously. The watch uploads the data to the database through smartphones and wireless communications. The database is always updated with the latest health data. A querier can send query to the database and obtain the statistic information about the user. However, the responses should not disclose any single data record to the querier. There are four steps for the query process.

\textbf{Step 1:} Query generation: querier sends the query sequence $Q$ to the database. The querier in this case could be the personal healthcare provider of the user or any other party (e.g. doctor, hospital, etc.) who needs to monitor the health conditions of the user. The sequence of queries can be average heart rate of the users for every 10 minute of her daily life over a week or month. This query can help the doctor to monitor the general pattern of changes in her patient's heart rate on a very fine-grained basis.
\newgeometry{left=0.63in,right= 0.63in,top=0.75in,bottom=1in}
\textbf{Step 2:} Aggregate query generation: based on the queries, the database records would be aggregated into data bins. For instance the database may contain user's heart rate for each minute but the querier (i.e. doctor) may only be interested in monitoring the information on a 5 minute basis. Thus the data bins generated by the database aggregates information of 5 records into a single bin.

\textbf{Step 3:} DP-compliant partitioning: the bins resulted from previous step will be then partitioned into a set of buckets based on the values of each bin, structure of the data and user defined threshold values indicating the level of granularity required by the querier. The purpose of this partitioning is to achieve more efficiency over answering the queries and also it is proven that coarser granularity achieved by a proper bucketing will increase the accuracy of the randomization~\cite{li2014data}. This process needs to use randomization which complies with differential privacy requirements because the structure of the buckets may reveal information and due to small changes in the database, private information in the database can be inferred. Also, because of general nature of aggregation and randomization, the patterns in the original data will usually be removed during this process. Our partitioning algorithm is proven to be differentially private and preserve the patterns of the original data.

\textbf{Step 4:} Randomizing Algorithm: the query results of buckets from previous step will be randomized by adding noise from Laplace distribution to preserve the privacy of the data. This step simply follows the data-independent Laplace mechanism which adds noise to the results regardless of the inputs. Then the results would be sent to the querier.

The first two steps are widely studied and the process suggested for them is fairly straightforward and standard, thus we skip further details about them and focus on the last two steps especially the third step as the main goal of this work.

\subsection{DP-Compliant Partitioning}
The partitioning of the data bins into the buckets has to be complying with differential privacy requirements because it has to be guaranteed that the bucketing is probabilistically the same for two neighboring databases, in other words, adding or removing a record from the database should be tolerated through a random variable. The process of partitioning itself follows a simple triple thresholded scheme:

\subsubsection{Algorithm setup and initialization}
The variable declarations and initializations should be done before the start of the algorithm process itself. The $Min, Max$ are non-negative numbers used to hold the values of maximum and minimum of bins in each bucket. Integers $i,j,size$ are used to hold the indexes of current bin, current bucket and size of current bucket respectively. Also $current$ is used to hold the value of previous bin (i.e. $x_{i-1}$). Three following threshold parameters are learned from public accessible information and are set based on user and querier's setup:
\begin{itemize}
	\item $T_D$: this value bound the maximum possible difference in maximum and minimum values of a single bucket to insure the uniformity of each bucket. (i.e. the bin values can dramatically change if this threshold is too large.)
	\item $T_L$: the maximum possible length of buckets is bounded by $T_L$ in order to evade creating over-sized buckets.
	\item $T_R$: this threshold which is naturally selected to be smaller than $T_D$, ensures that changes in the data are observed and differentiated from changes which generally will be captured by $T_D$, because the change in two adjacent bins may actually be smaller than the $T_D$ threshold but since it has happened in a very small period of time, it has important information in it and has to be preserved and reflected for the doctor. For instance increasing heart rate during daily activities will be distinguished from changes during workout.
\end{itemize}
Due to the privacy requirement of the partitioning algorithm, it is necessary to randomize $T_D$ and $T_R$ threshold parameters. Thus, using the Laplace mechanism, with privacy budget $\epsilon_1$ we produce random noises $Y,Y'$ which will be added to $T_D$ and $T_R$ and form $\hat{T}_D$ to $\hat{T}_R$ respectively.

\subsubsection{Partitioning process}
The partitioning process is done using an efficient and simple process of scanning from beginning of the data domain to the end (i.e. first to the last bin) with possible single backtracks during scan. The process starts with placing the first bin into the fist bucket and continues to next bins. In case uniformity of the bins complies with threshold requirements, the bin would be added to the same bucket, otherwise a new bucket would be created and process continues with the new bucket. The first condition to check is the $T_R$ threshold because of its smaller value (line 15). In case of this condition not satisfying, two single bin buckets (each containing one of the unbalanced bins) are required to be created. Based on the size of the current bucket, three different cases are considerable (lines 16, 27 and 31). Furthermore, after this condition the two remaining thresholds would be examined (line 38) and based on that condition, either a new bucket would be created or the current bucket would be enlarged.
Creating new bucket is simply done by pushing the current bucket (i.e. $b_j$) into the result set (i.e. $B$) and incrementing $j$ followed by resetting $size$ to $0$.

%\vspace{-0.2cm}

\begin{algorithm}
	\small
	\caption{Private Partitioning Algorithm}\label{part}
	\begin{algorithmic}[1]
		\item \textbf{Input:} database $D$, $T_D$, $T_L$, $T_R$, $\epsilon_1$\\
		\textbf{Output:} a set of histogram buckets $B$\\
		\textit{Initialization}: Set $size = 0; i = 1; j = 1$, $B = \emptyset$,\\$\hat{T}_D=T_D+Y, \hat{T}_R=T_R+Y'$		\Comment{$Y,Y' \AC Lap(1/\epsilon_1)$}
		\While{$i \leq length(D)$}
		\State $current=NULL;$
		\If{$(size == 0)$}
		\State $b_j \leftarrow x_i$ \Comment{Put bin $x_i$ into bucket $b_j$}
		\State $Min= Max=current=x_i;$
		\State $current = x_i;size++;i++;$ \Comment{Goto next bin}
		\EndIf
		\State $Max = max(Max,x_i);Min = min(Min,x_i)$
		\If{($current \neq NULL$ and $|current - x_i| > \hat{T}_R$)}
		\If{$size == 0$}
		\If{($B[-1].length > 1$)} %\Comment{Previous bucket has more than one bin, has to be partitioned}
		\State $last = B.pop();b_j =last.pop()$ %\Comment{Put last bin of previous buket into a new bucket}
		\State $B \leftarrow last$; $B \leftarrow b_j$; $j++$; %\Comment{Putting both buckets into the results and going to next bucket}
		\State $b_j \leftarrow x_i$; $B \leftarrow b_j$; %\Comment{Creating second single bin bucket and putting it into the results}
		\State  $j++; current = x_i; i++; size = 0; $
		\Else \Comment{Last bucket is already single bin}
		\State $b_j \leftarrow x_i; B \leftarrow b_j$
		\State  $j++; current = x_i; i++; size = 0; $
		\EndIf
		\ElsIf{$size ==1$}
		\State $B \leftarrow b_j; j++;b_j \leftarrow x_i; B \leftarrow b_j; j++;$
		\State $current = x_i; size = 0; i++$
		\ElsIf{$size >1$}
		\State $last = b_j.pop();B\leftarrow b_j; j++;$
		\State $b_j \leftarrow last; B \Leftarrow b_j; j++;$
		\State $b_j \leftarrow x_i; B \leftarrow b_j; j++;$
		\State $current = x_i; size = 0; i++$
		\EndIf \Comment{$size$ check}
		\EndIf \Comment{$\hat{T}_R$ check}
		\If{$((Max-Min \leq \hat{T}_D)$ and $(size \leq T_L)$}
		\State $b_j \leftarrow x_i$ \Comment{Put bin $x_i$ into bucket $b_j$}
		\State $current = x_i;size++;i++;$ \Comment{Goto next bin}
		\Else
		\State $B \leftarrow b_j$ \Comment{Done with this bucket}
		\State $current = x_i;size=0;j++;$\Comment{Goto next bucket}
		\EndIf
		\EndWhile\label{euclidendwhile}
		\State \textbf{return} $B$\Comment{Set of buckets}
	\end{algorithmic}
\end{algorithm}

\subsection{Randomizing Algorithm}

After partitioning of the data into buckets, adding noise to the average value of bins in each bucket would simply satisfy DP as long as the added noise is generated with proper setting and noise scale. In this step we use the Laplace distribution for generating noise. Considering the maximum possible change in values of bins in neighboring databases is $\alpha$, since value of each bucket is the average value of its bins and having buckets with the $size =1$ is considerable, the Laplace scale for this step has to be $b=\frac{\alpha}{\epsilon}$ for achieving $\epsilon-$DP.

\section{proof of privacy}

This part shows the proof of privacy for Algorithm 1. Let $d_0, d_1$ be neighboring databases and $\mathcal{A}(d_0), \mathcal{A}(d_1)$ be the output of the algorithm on these databases; $Max_{ik}$, and $Min_{ik}$ be the maximum and minimum value of bins in $ith$ bucket $b_i$ of $d_k$. $Y$ and $Y'$ are Laplace random variables and $f_y, f_{y'}$ are their density function, therefor $\hat{T}_D=T_D+Y$ and $\hat{T}_R=T_R+Y'$ are the randomized threshold values. To show that the algorithm is $\epsilon-$differentially private, we need to show that result of partitioning of $d_0$ and $d_1$ are probabilistically equivalent, thus it is sufficient to prove: $Pr(\mathcal{A}(d_0) = B) \leq
e^{\epsilon} \times Pr(\mathcal{A}(d_1) = B)$. The parameters effective in partitioning results are the threshold values (i.e. $T_L$, $\hat{T}_D$ and $\hat{T}_R$), but $T_L$ is not a random variable, therefor only $\hat{T}_D$ and $\hat{T}_R$ are effective in randomness of the result. Suppose the maximum difference in value of bins in two neighboring databases is bounded by $\alpha$. This $\alpha$ can be learned from public accessible data based on the type of queries and maximum possible values for the under study filed of the record. For each bucket, we have $Max_i - Min_i < \hat{T}_D$ and $|current - x_i| < \hat{T}_R $. These two inequalities can be generalized to the whole database:

\begin{small}
	\[\frac{Pr(\mathcal{A}(d_0) = B)}{Pr(\mathcal{A}(d_1) = B)} \leq e^{\epsilon} \Leftrightarrow \mathcal{X}=(\frac{\prod_{b_i \in d_0} Pr(Max_{i0} - Min_{i0} < \hat{T}_D) }{\prod_{b_i \in d_1} Pr(Max_{i1} - Min_{i1} < \hat{T}_D)}\]\[\times \frac{ \prod_{b_i \in d_0} Pr(|current - x_{i0}| < \hat{T}_R)}{ \prod_{b_i \in d_1} Pr(|current - x_{i1}| < \hat{T}_R)}) \leq e^{\epsilon}\]
\end{small}

Using the \textit{sequential composition} property of DP, taking $\epsilon = \epsilon_1 + \epsilon_2$, we have:

\begin{small}
	\[\mathcal{X} = (\frac{\prod_{b_i \in d_0} Pr(Max_{i0} - Min_{i0} < \hat{T}_D) }{\prod_{b_i \in d_1} Pr(Max_{i1} - Min_{i1} < \hat{T}_D)} \leq e^{\epsilon_1})\] \[\ \times(\frac{ \prod_{b_i \in d_0} Pr(|current - x_{i0}| < \hat{T}_R)}{ \prod_{b_i \in d_1} Pr(|current - x_{i1}| < \hat{T}_R)}) \leq e^{\epsilon_2})\]
\end{small}

Thus if both parts of this equation holds, the algorithm is $\epsilon-$differentially private. We try to solve these inequalities (i.e. $\mathcal{X}_1$ and $\mathcal{X}_2$) in order to find the required Laplace distribution scale for satisfying $\epsilon-$DP. Suppose the changed record falls into bucket $b_i$ ($ith$ bucket of $d_k$). Altering $d_0$ or $d_1$ are equivalent, thus we only consider one case. For the first inequality (i.e. $\mathcal{X}_1$), if the changed record is between minimum and maximum values of the bucket (i.e. $Min_{i0} \leq x_{i0} \leq Max_{i0}$), then the $Max_{i0}$ and $Min_{i0}$ will not change and $Pr(Max_{i0} - Min_{i0} < \hat{T}_D) = 1 \leq e^{\epsilon_1}$ for every $\epsilon_1$. But if the changed value effects either $Max_{i0}$ or $Min_{i0}$, we need to find the suitable Laplace scale ($b=s/\epsilon_1$) in order to have this change tolerated. The $Max_{i0}$ and $Min_{i0}$ are changes by $\alpha$ units and to the value of the $\mathcal{X}_1$, increasing $Min_{i0}$ is equivalent to decreasing $Max_{i0}$ and vice versa, thus we only consider change of $Max_{i0}$. We take $Y \AC Lap(s/\epsilon_1), t=Max_{i0}-Min_{i0}$ and $u=t-T_D$, then we consider two cases of changing $Max_{i0}$:

1) If $Max_{i0} \leftarrow Max_{i0} + \alpha$:
\begin{small}
	\[\mathcal{X}_1=\frac{Pr(t+\alpha < \hat{T}_D)}{Pr(t<\hat{T}_D)} = \frac{Pr(Y > u+\alpha)}{Pr(Z>u)} <1\leq e^{\epsilon_1} \]
\end{small}
2) If $Max_{i0} \leftarrow Max_{i0} -\alpha$:
\begin{small}
	\[\mathcal{X}_1=\frac{Pr(t-\alpha < \hat{T}_D)}{Pr(t<\hat{T}_D)} = \frac{Pr(Y > u-\alpha)}{Pr(Z>u)} = \frac {\int_{u-\alpha}^{+\infty}f_y(y)dy}{\int_{u}^{+\infty}f_y(y)dy} \leq e^{\epsilon_1}\]
\end{small}
We consider following cases in order to solve the above inequality:\\
\begin{small}
	$\bullet$ $u \geq \alpha: \mathcal{X}_1 = e^{\epsilon_1/s} \leq e^{\epsilon_1} \Rightarrow \epsilon_1/s \leq \epsilon_1 \Rightarrow 1/s \leq 1 \Rightarrow s \geq 1 $ \\
	$\bullet$ $0 < u <\alpha:$
	\[\mathcal{X}_1=\frac {\frac{1}{2} + \int_{u-\alpha}^{+\infty}f_y(y)dy}{\int_{u}^{+\infty}f_y(y)dy} = \frac{2-e^{\frac{u-\alpha}{b}}}{e^{\frac{-u}{b}}} \leq e^{\epsilon_1}
	\]
\end{small}
Taking $v = e^{{\frac{u}{b}}}$, then $\mathcal{X}_1=2v-e^{\frac{-\alpha}{b}}v^2 \leq e^{\epsilon_1} \Rightarrow s \geq \alpha$\\
\begin{small}
	$\bullet$ $u \leq 0:$
	\[\mathcal{X}_1=\frac {\frac{1}{2} + \int_{u-\alpha}^0f_y(y)dy}{\frac{1}{2} +\int_{u}^0f_y(y)dy} = \frac{2-e^{\frac{u-\alpha}{b}}}{2-e^{\frac{u}{b}}} \leq e^{\epsilon_1}\]
	\[\Leftrightarrow e^{\epsilon_1}(e^{u\epsilon_1})^{\frac{1}{s}} - [e^{(u-\alpha)\epsilon_1}]^\frac{1}{s} \leq 2e^{\epsilon_1}-2\]
\end{small}
Taking $s=\alpha$, the inequality above holds. Thus for the first part of the equation (i.e. $\mathcal{X}_1$), the Laplace scale $b =  \frac{\alpha}{\epsilon_1}$ is sufficient for DP.

For the second part (i.e. $\mathcal{X}_2$), we have to make sure the difference between changed record and its next (and previous) record will be tolerated through $\hat{T}_R$ and will not result in changing the buckets. Considering the differences between values of consecutive bins, increasing the larger value is equivalent to decreasing the smaller value and vice versa, thus considering only changes in one value (i.e. $x_{i0} \leftarrow x_{i0}+\alpha$ and $x_{i0} \leftarrow x_{i0}-\alpha$) is sufficient for the proof. We have:

\begin{small}
	\[\mathcal{X}_2=\frac{Pr(|x_{(i-1)0} - x_{i0}| < \hat{T}_R)}{Pr(|x_{(i-1)1} - x_{i1}| < \hat{T}_R)} \times \frac{Pr(|x_{(i+1)0} - x_{i0}| < \hat{T}_R)}{Pr(|x_{(i+1)1} - x_{i1}| < \hat{T}_R)}\]
\end{small}

Also, relative change between $x_{(i-1)1}$ and $x_{i1}$ is equivalent to change between $x_{(i+1)1}$ and $x_{i1}$, thus only considering one case is sufficient for the proof. Considering the large value in the adjacent bins as $Max_{i0}$ and the smaller vales as $Min_{oi}$ through the exact same argument as the argument for previous part, the proof of privacy for $s = \alpha \Rightarrow b = \frac{\alpha}{\epsilon_2}$ is immediate.

\section{evaluation}
In this section, we evaluate the performance of our newly-designed algorithm. We have conducted real experiments on captured heart rates from wearable devices attached to a user during two weeks. Heart rate usually ranges between 50 and 210, and thus we choose $\alpha=\frac{160}{14}$ as the sensitivity. %We draw Laplace noises with Laplace scale $\alpha$, thus $\epsilon_1=\epsilon_2=1$ and the whole mechanism is $(\epsilon_1+\epsilon_2)-$differentially private.
Obviously, for a bigger time period of data collection, the required Laplace scale for achieving same privacy budget gets smaller as a result of less sensitivity due to aggregation over more data.
The experiments are run on a Linux machine with an Intel Core i7 3.5 GHz processor and 8 GB of DDR3 memory.
\subsection{Pattern preservation evaluation}
\begin{figure}[htbp]
	\centering
	\includegraphics[width=.48\textwidth]{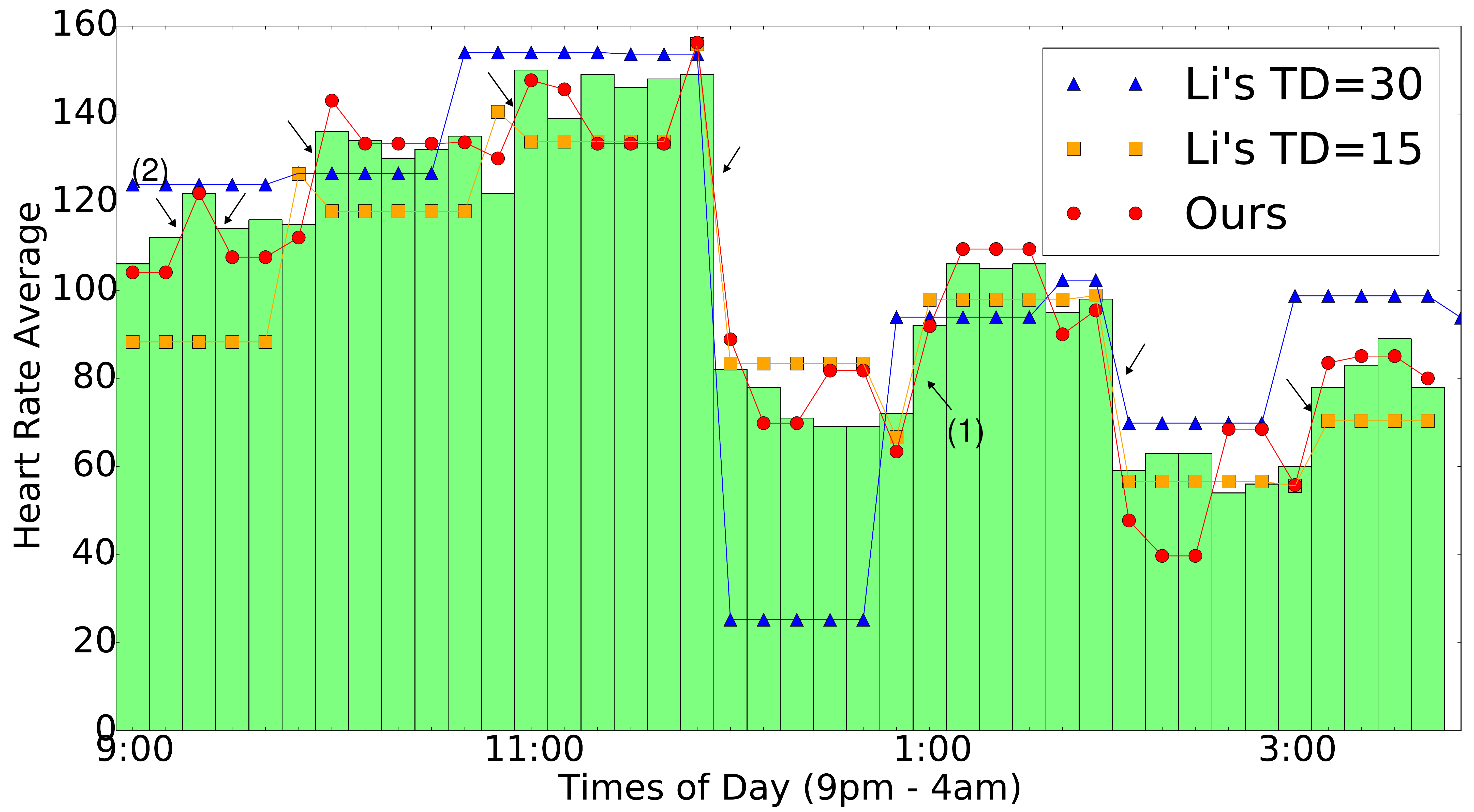}
	\vspace{-0.3cm}	
	\caption{Pattern preservation results based on 10 minutes query sequence}\label{intuitive}
	\vspace{-0.4cm}	
\end{figure}

We compare our algorithm and the state-of-the-art proposed by Li el al.~\cite{li2015ehealth} in terms of capturing the patterns of the original data and reflecting them in the final output. To do so we define pattern preservation percentage as $\dfrac{detected\_rapid\_changes}{all\_rapid\_changes\_in\_data} \times 100$ Our algorithm uses an additional threshold parameter $T_R$ as maximum difference allowed for two adjacent bins to be combined into one bucket. In case this threshold is violated, our algorithm divides the two bins into two single-bin buckets in order to emphasis on this rapid change as opposed to Li's algorithm which does not capture rapid changes. In our designed experiment we used $T_L = 4, T_D = 30$ and $T_R=15$ for our algorithm and for Li's algorithm we used two settings for their $T_D$ to be as small as our $T_R$ and also same value as our $T_D$, and have assigned the same value for their $T_L$ as for ours.
We tested both algorithms on a set of collected heart rate per minute data stored for two weeks. Fig. \ref{intuitive} is a sample result of partitioning and randomizing of both algorithms. The black arrows are rapid changes which are captured by our algorithm. The arrow marked by (1) is an example of failing Li's algorithm with $T_D=30$ in capturing the rapid change and (2) is an example in which both Li's algorithms fail in doing so.
As shown in Fig. \ref{percentile} (a) our algorithm does better than Li's in preserving the pattern of original data with the same $T_D$ threshold. In case of selecting Li's $T_D$ to be as small as our $T_R$, obviously every rapid change in data is also captured as subset of $Max-Min$ factor which Li's algorithm is focused on, thus almost (because of the randomization effect) equal preservation percent is reasonable in this case.

The very important point is that our algorithm other than detecting the rapid change, using the idea of creating single bin buckets at the places of rapid change is preserving the change from being smoothed out through the averaging in the bucket. Specifically, as shown in Fig. \ref{percentile}, our algorithm outperforms Li's (with the same $T_D$ values) in percentage of preserved rapid changes with $70.81 \%$ compared to $40.1 \%$. As mentioned Li's algorithm with $T_D=15$ is able to perform almost equal to ours ($68.36\% $) but may lose the found pattern by averaging bin values over their bucket. The values are computed by averaging over results of 1000 experiments.

\subsection{Error analysis}

\begin{figure*}[htbp]
	\centering
	\includegraphics[width=.29\textwidth]{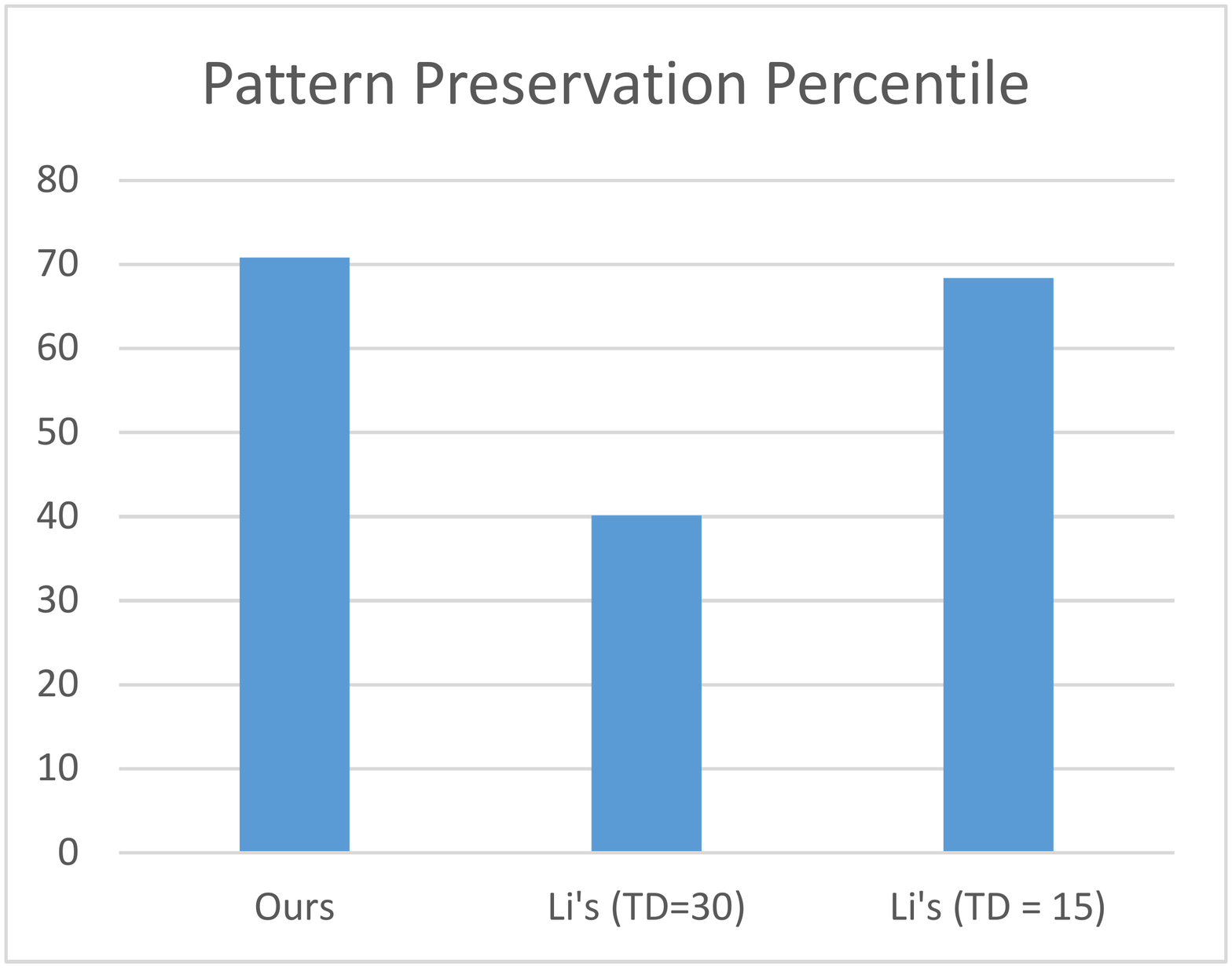}
	\includegraphics[width=.29\textwidth]{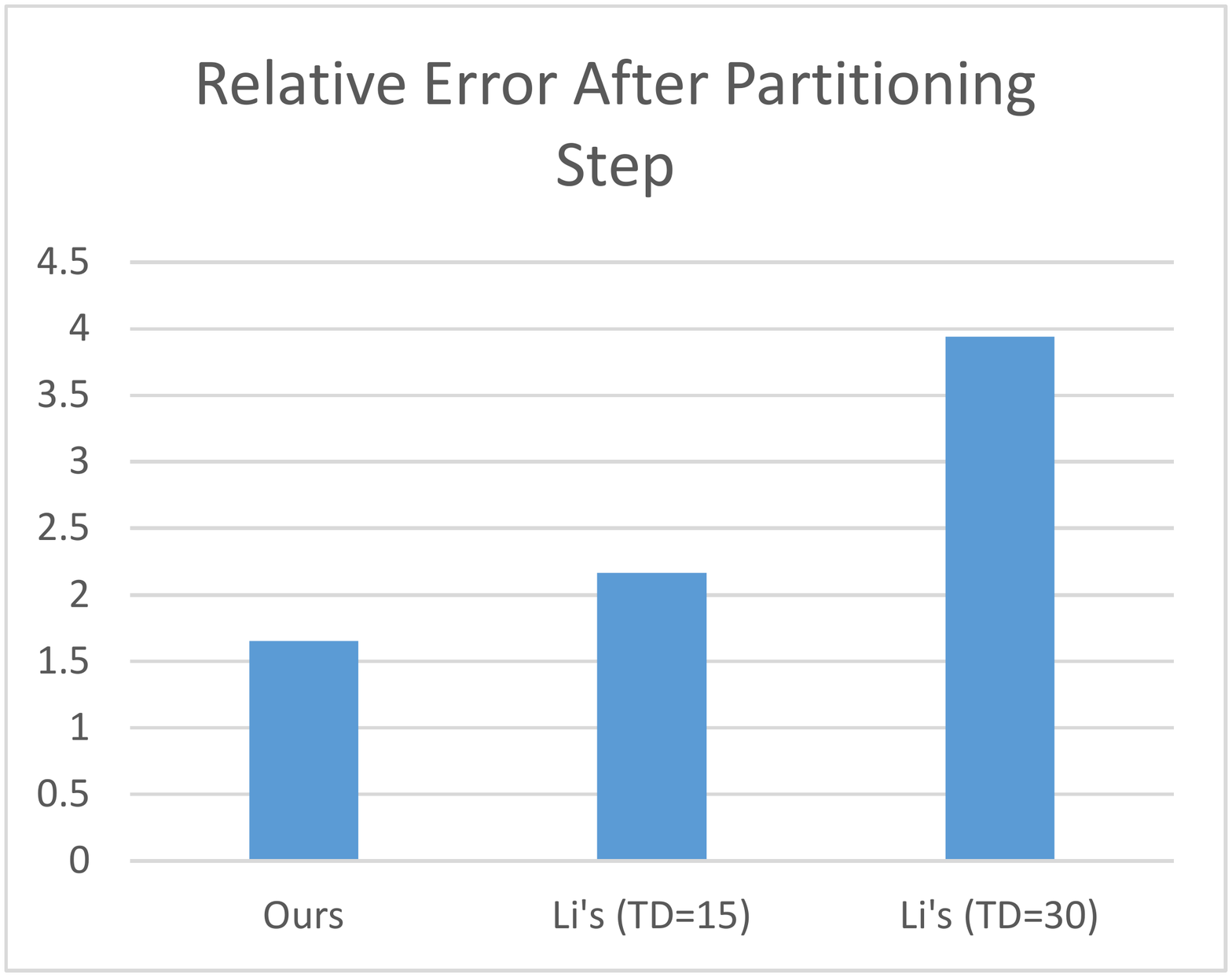}
	\includegraphics[width=.29\textwidth]{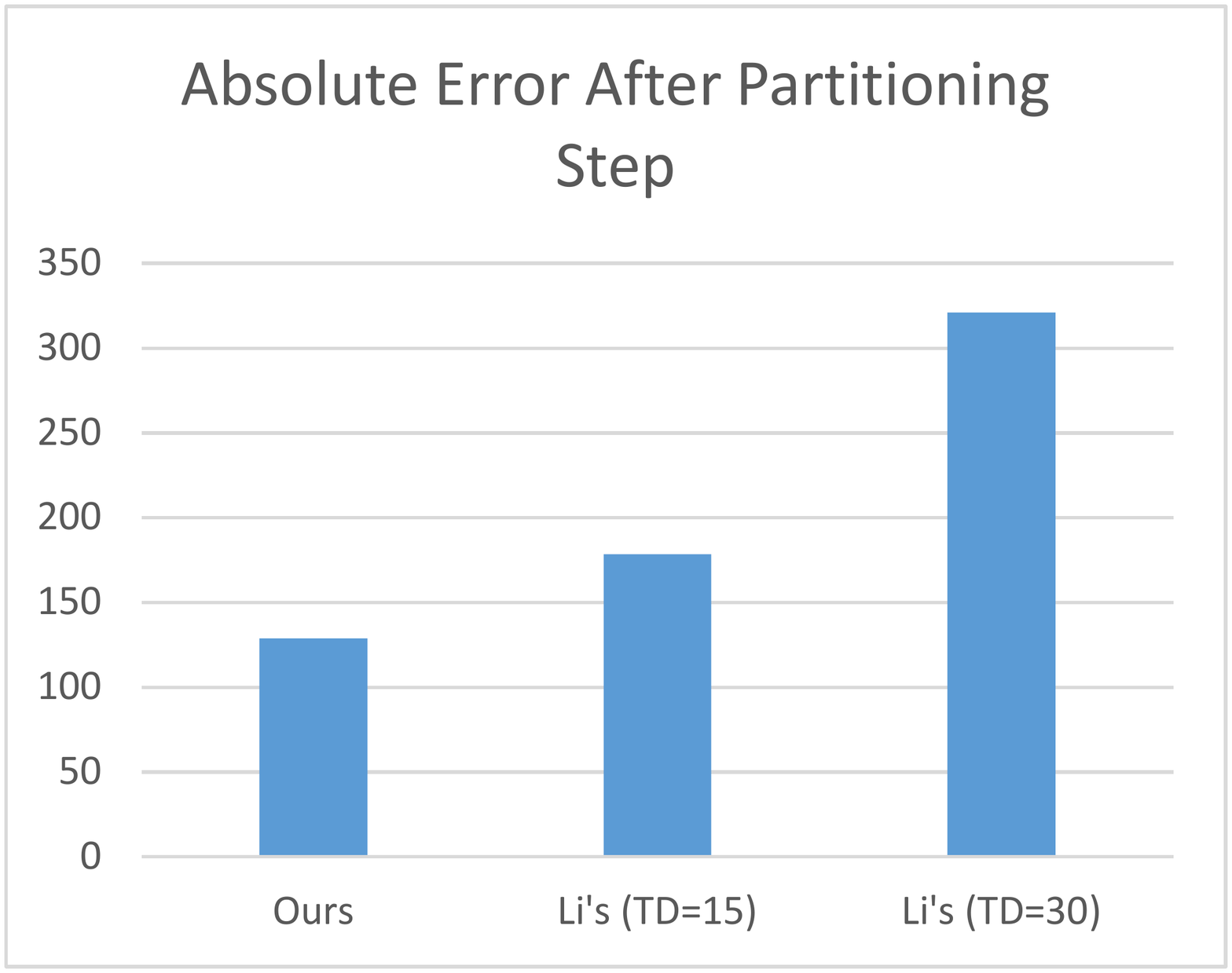}
	\vspace{-0.3cm}
	\caption{(a) Pattern preservation percentile. (b) \& (c) Relative and absolute error comparison comparison}\label{percentile}
	\vspace{-0.6cm}
\end{figure*}

In this subsection we compare our algorithm with Li's in terms of absolute and relative error values in two steps and argue that preserving the pattern will result in higher accuracy (less error) because of the structural benefit gained in our algorithm. The reason is better uniformity between bins in a bucket which makes the averaging closest to the actual values. Obviously failing in dividing the unbalanced bin from others will result in higher error in averaging and leads to less accuracy at the end. It is important to mention that in the later steps of the scheme because of added significant Laplace noise (compared to error values) drawn from the same scale in both ours and Li's algorithms, only the errors in partitioning step are comparable in two algorithms. As shown in Fig. \ref{percentile} (b, c) our algorithm outperforms Li's  in both absolute and relative error metrics at partitioning step.

\subsection{Time complexity}
Due to the higher complexity of our algorithm than Li's, also considering the efficiency of their algorithm which does the partitioning in only one scan from the beginning to the end of the data (i.e. $O(n)$), a marginal increase in time complexity is acceptable as a trade-off for more accurate pattern preservation. Specifically our algorithm also scans the data only once, thus our algorithm has the same time complexity at Li's. The only overhead is in case of the $T_R$ threshold violation in which there are several cases considerable and more computations necessary to handle the case. As average over 1000 experiments, Li's times for only partitioning and both steps are $0.734$ ms and $1.012$ ms and ours does it on $0.763$ ms and $1.09$ ms respectively.

\section{Related works}

Security and privacy in eHealth data has been widely studied in the literature~\cite{hersh2002medical, dong2011challenges, sahama2013security} because of the private and sensitive nature of the date dealt with. The mechanism of mHealth~\cite{istepanian2006m} is a more recent extension to eHealth for revolutionizing the healthcare systems though mobile communications and has been highly attracted lately~\cite{kay2011mhealth} and all privacy concerns raised in eHealth are applicable to mHelath. Our work is specifically focused on privacy preservation in mHealth data and considers specific requirements of it due to the sensitive nature of its data. While standard methods of secrecy hide the content of the message, covert communication in wireless environments~\cite{soltani2014covert, soltani2018covert} and computer networks~\cite{soltani2015covert, soltani2016covert} hides the existence of the communication.

Differential Privacy ~\cite{DBLP:reference/crypt/Dwork11,dwork2006calibrating,dinur2003revealing} is a newly emerged mechanism for private data publication with strong mathematical proof of privacy preservation. Dwork et al.~\cite{dwork2006calibrating} introduced the Laplace mechanism (LM) for generating noise, which is commonly used in the literature. LM uses a notion of sensitivity of the query ($\Delta$) for finding the proper Laplace scale ($b$). In this work we use the same method of sensitivity and Laplace scale.%Data release is classified as interactive and non-interactive \cite{blum2005practical} models.
Private partitioning of histograms under differential privacy has been widely studied. Blum et al.~\cite{blum2013learning} have introduced a one-dimensional histogram, while Xiao et al.~\cite{xiao2010differentially} have suggested a multi-dimensional one using a wavelet-based technique. Xu et al.~\cite{xu2013differentially} have classified and compared different existing approaches for histogram publishing that takes the structure of the bins into account. They have proposed two different algorithms for publication and have stated that the granularity of the partitioning has impact on accuracy of the results. A new partitioning algorithm achieving higher accuracy has been introduced in \cite{li2014data} by C. Li et al. and the performance of their method in partitioning and update is further improved by H. Li et al. in \cite{li2015ehealth} but neither these works consider capturing the slight and rapid changes patterns in the original data which is main goal of our work.

\section{Conclusion}
In this paper, we proposed a novel private partitioning scheme for mHealth data under DP that preserves the patterns of the original data and reflects it to the results. We gave strict privacy proof to show this scheme is differentially private. Also through evaluation in real experiment we showed that the pattern of the original data is mostly reflected into the final results as opposed to the current state-of-the-art which is not able to properly detect and preserve the pattern. We showed our algorithm gains more accuracy due to its benefit from pattern preservation.
In our future work, we will explore more characteristics of mHealth data in order to gain more utility for the querier, and specifically we will focus on different methods for improving the accuracy of pattern preservation and the results through relative error.

\bibliographystyle{IEEEtran}
\bibliography{ref}

\end{document}